 \documentclass[pmlr,twocolumn]{jmlr} 

\usepackage[utf8]{inputenc} 
\usepackage[T1]{fontenc}    
\usepackage{hyperref}       
\usepackage{url}            
\usepackage{booktabs}       
\usepackage{amsfonts}       
\usepackage{nicefrac}       
\usepackage{microtype}      
\newcommand{\E}{\ensuremath{\mathbb{E}}} 
\newcommand\pois{\text{Pois}}
\newcommand\bin{\text{Binom}}
\usepackage{placeins} 
\usepackage{graphicx}
\usepackage{amsmath}

\usepackage{rotating}

\usepackage{booktabs}
\usepackage[load-configurations=version-1]{siunitx} 

\theorembodyfont{\upshape}
\theoremheaderfont{\scshape}
\theorempostheader{:}
\theoremsep{\newline}

 \jmlrvolume{ML4H Extended Abstract Arxiv Index}
\jmlryear{2020}
\jmlrsubmitted{2020}
\jmlrpublished{}
\jmlrworkshop{Machine Learning for Health (ML4H) 2020}

\title[Assessing racial inequality in COVID-19 testing with Bayesian threshold tests]{Assessing racial inequality in COVID-19 testing\titlebreak with Bayesian threshold tests}

\author{%
\Name{Emma Pierson} \Email{emmap1@cs.stanford.edu} \\
\addr  Microsoft Research, Cambridge, MA, USA
}

\makeatletter
\def\set@curr@file#1{\def\@curr@file{#1}} 
\makeatother
\usepackage[load-configurations=version-1]{siunitx} 

\begin{document}

\maketitle

\begin{abstract}
There are racial disparities in the COVID-19 test positivity rate, suggesting that minorities may be under-tested. Here, drawing on the literature on statistically assessing racial disparities in policing, we 1) illuminate a statistical flaw, known as \emph{infra-marginality}, in using the positivity rate as a metric for assessing racial disparities in under-testing; 2) develop a new type of Bayesian \emph{threshold test} to measure disparities in COVID-19 testing and 3) apply the test to measure racial disparities in testing thresholds in a real-world COVID-19 dataset.
\end{abstract}

\section{Introduction} 

A widely used metric in monitoring COVID-19 outbreaks is the \emph{positivity rate}, defined as the fraction of COVID-19 tests which are positive~\citep{jhu_positivity_rate,who_positivity_rate}. A low positivity rate suggests that an area has enough testing to properly monitor its outbreak; a high positivity rate suggests under-testing. 

In the United States, there are large racial disparities in COVID-19 cases and deaths per capita~\citep{new_york_times_racial_inequity}, prompting recommendations that the positivity rate be reported broken down by race~\citep{positivity_rate_by_race_2,positivity_rate_by_race_1}. While this data is not yet systematically available on a national level, the data that is available reveals large racial/ethnic disparities (Figure \ref{fig:positivity_rate_by_state}), consistent with prior work~\citep{martinez2020sars,cordes2020spatial,bilal2020spatial}. The Black positivity rate is higher in all states with data than the white positivity rate; similarly, the Hispanic positivity rate is higher in all states than the non-Hispanic positivity rate, suggesting under-testing of Black and Hispanic populations.

Motivated by these racial disparities in the positivity rate, in this work we make three contributions. First, we illuminate a statistical flaw, known as \emph{infra-marginality}, in using the positivity rate as a metric for assessing racial disparities in under-testing, drawing on the literature on measuring racial disparities in policing. Second, we describe how a Bayesian \emph{threshold test} approach, which has been used to measure racial disparities in policing, can be used to measure disparities in COVID-19 testing, and develop a version of the test suitable for the COVID-19 setting. Third, we use the test to measure racial disparities in testing thresholds in a real-world COVID-19 dataset. We conclude by discussing broader applications of threshold tests in medicine.

\begin{figure}[htbp]
\floatconts
  {fig:positivity_rate_by_state}
  {\caption{Positivity rates in 6 states broken down by race and ethnicity reveals higher positivity rates for non-white populations. Data from \citet{covidtrackingproject}.}}
  {\includegraphics[width=1.1\linewidth]{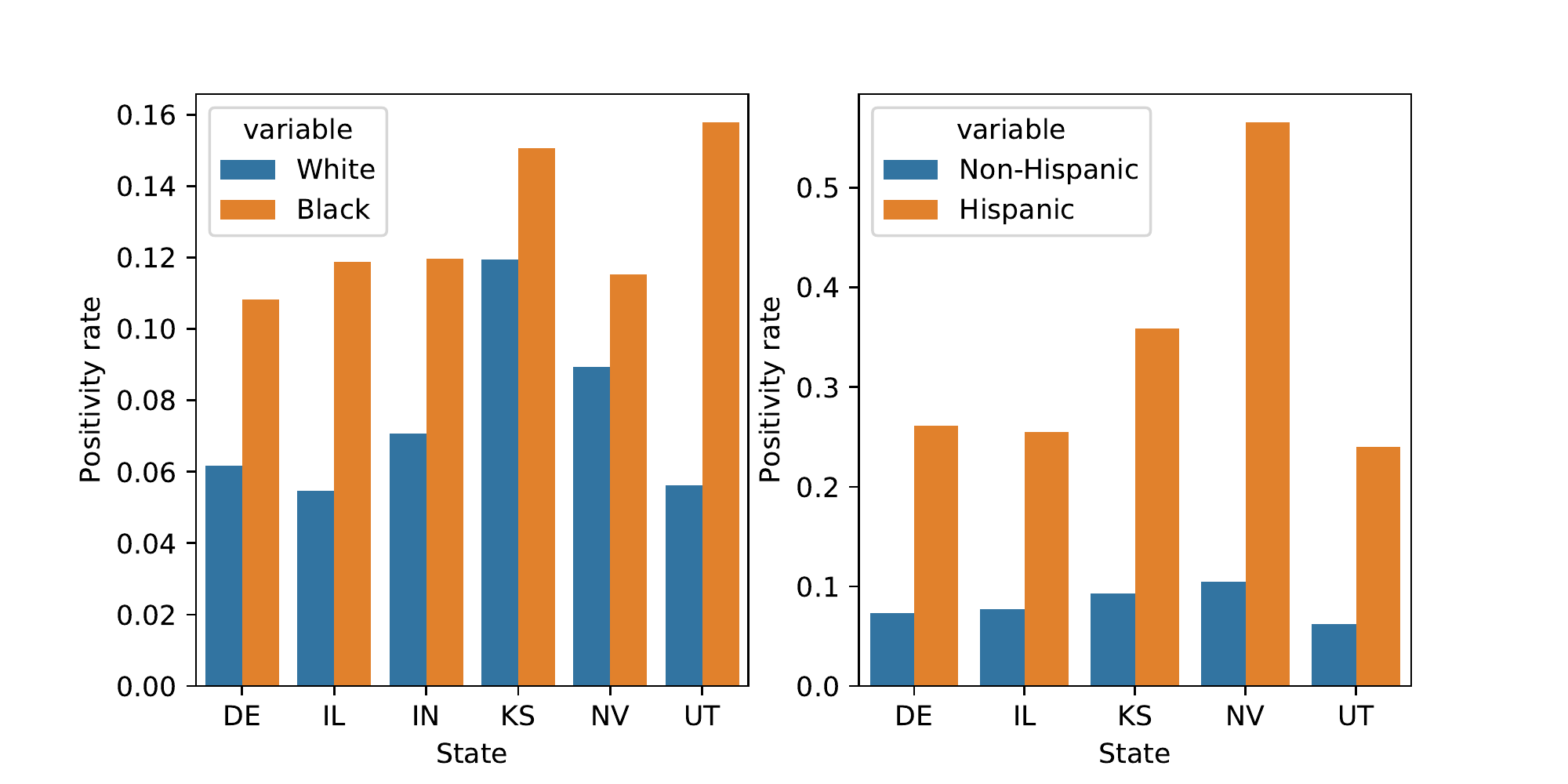}}
\end{figure}

\section{Infra-marginality} 

Breaking down the COVID-19 positivity rate by race is an example of an ``outcome test'', a widely used technique for measuring racial bias in decision-making by examining the \emph{outcomes} of decisions~\citep{ayres2002outcome,becker1993nobel,carr1993federal}. Outcome tests have been applied in diverse domains from policing to lending. 
In policing, a frequently used outcome is whether a police search finds contraband: if searches of white drivers find contraband 90\% of the time, but searches of Black drivers find contraband only 10\% of the time, it suggests that the police are searching white drivers only when they’re very likely to be carrying contraband, but searching Black drivers at a lower threshold of evidence. In lending, the outcome is whether someone granted a loan repays it: if white loan recipients pay back loans only 10\% of the time, but Black loan recipients pay back loans 90\% of the time, it suggests that Black applicants are receiving loans only when they are very likely to pay them back, at a higher threshold of evidence. In COVID-19 testing, the outcome is whether a COVID-19 test comes back positive; if tests of Black patients find COVID-19 90\% of the time, but tests of white patients find COVID-19 only 10\% of the time, it suggests that Black patients are being tested only when they are much more likely to have COVID-19, at a higher threshold of evidence. In all these examples, then, racial differences in outcomes are concerning because they suggest racial differences in the \emph{probability thresholds} people face --- to undergo a police search, to receive a loan, or to get a COVID-19 test. 

But the previous literature on outcome tests also illuminates a problem with simply examining the positivity rate, called \emph{infra-marginality}~\citep{ayres2002outcome}, which we explain by adapting an example from ~\citet{simoiu2017problem}. Imagine that there are two races --- white and Black --- and within each race there are two equally-sized groups --- one who is very unlikely to have COVID-19, and one who is quite likely. Imagine these groups are easy to tell apart --- one group is showing COVID-19 symptoms, for example, and one group is asymptomatic. 5\% of the asymptomatic patients have COVID-19, regardless of their race. 50\% of the white symptomatic patients have COVID-19, and 75\% of the Black symptomatic patients have COVID-19. Finally, imagine there is no racial bias in who is tested: everyone who is more than 10\% likely to have COVID-19 is tested, regardless of race, so the same probability threshold is applied to both races. All symptomatic patients will be tested, producing a positivity rate of 50\% for white patients and 75\% for Black patients. We will incorrectly conclude from the higher positivity rate among Black patients that they are being under-tested relative to whites --- that is, tested only when they are more likely to have COVID-19. But in fact, in this hypothetical, everyone faces the same 10\% testing threshold. We reach this misleading conclusion because the statistic we’re measuring --- the positivity rate --- is not the same as the \emph{probability threshold} at which patients are being tested. (We note that positivity rate analysis can also yield a misleading result in the \emph{opposite} direction, where it fails to show racial disparities even though there \emph{are} disparities in testing thresholds.) 

In general, if two race groups have very different \emph{risk distributions} (in the hypothetical example above, the Black risk distribution is right-shifted) simply looking at the positivity rate may yield misleading conclusions. Figure \ref{fig:hypothetical_risk_distributions} illustrates this graphically for continuous risk distributions. In the case of COVID-19, infra-marginality is not a hypothetical concern: per capita infection rates are much higher in Black populations than white populations, so it is plausible that there might be dramatic differences in the risk distributions. 

\begin{figure}[htbp]
\floatconts
  {fig:hypothetical_risk_distributions}
  {\caption{Hypothetical example illustrating that if two racial groups (red and blue lines) have very different distributions of COVID-19 risk, the same testing threshold produces different distributions above the threshold, and therefore different positivity rates.}}
  {\includegraphics[width=1\linewidth]{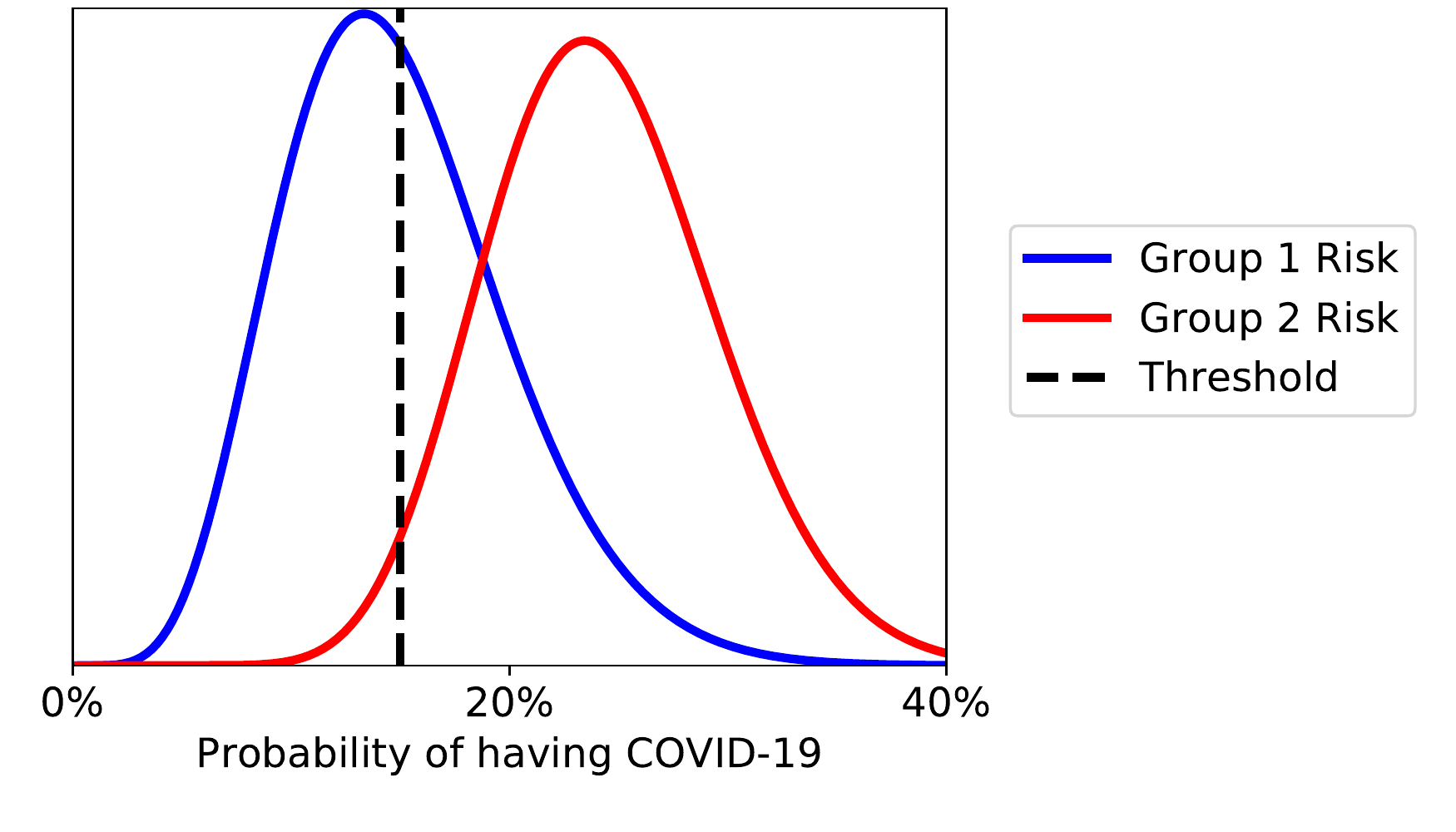}}
\end{figure}

This threshold which people are tested is hard to measure --- unlike the positivity rate, it's not a simple fraction directly observable from the data, but a latent quantity that must be inferred. \emph{Threshold tests} attempt to infer both the thresholds and risk distributions and thereby circumvent the problem of infra-marginality. 

\section{Threshold tests} 

Threshold tests for policing, proposed in~\citet{simoiu2017problem} and applied in ~\citet{pierson2018fast,pierson2020large}, use a Bayesian model to simultaneously infer the race- and location-specific risk distributions and thresholds. For brevity, we refer the reader to~\citet{simoiu2017problem} for a description of the original threshold test in the context of policing, and here describe only our adapted generative model for the COVID-19 context; Appendix~\ref{sec:further_model_details} details further how the two models differ.

\paragraph{Observed data.} We assume that we observe three pieces of information for all race groups $r$ and counties $d$: the population of the race group in the county, $n_{rd}$; the cumulative number of COVID-19 tests for the race group in the county $t_{rd}$; and the cumulative number of COVID-19 cases (positive tests) for the race group in the county, $c_{rd}$. 

\paragraph{Generative model.} On each day, the probability $p$ that a person of race $r$ in county $d$ has COVID-19 is drawn from a race and county-specific \emph{risk distribution} – a probability distribution on $[0, 1]$. $p$ represents the probability a person has COVID-19 given their relevant observable characteristics - for example, whether they are coughing and whether they have recently been to large gatherings. Each person gets tested if $p$ exceeds a race and county-specific testing threshold, $z_{rd}$. 

We let $P_{rd}$ denote the random variable corresponding to the risk distribution for each race and county. The probability a person of a given race in a given county will get tested, $f_{rd}$, is the proportion of the risk distribution that lies above the threshold, $p(P_{rd} > z_{rd})$, that is, the complementary cumulative distribution function of the risk distribution.  The probability a test will be positive, $g_{rd}$, is the expected value of the risk distribution conditional on being above the threshold: $\E(P_{rd} | P_{rd} > z_{rd})$. The observed data are drawn as as follows: 

\begin{align}
    t_{rd} &\sim \pois(n_{rd} \cdot f_{rd})\\
    c_{rd} &\sim \bin(t_{rd}, g_{rd})
\end{align}

The latent parameters of the model are the thresholds $z_{rd}$ and the parameters of the risk distributions. Following~\citet{pierson2018fast,pierson2020large}, we parameterize the risk distributions as discriminant distributions, which are two-parameter distributions on $[0, 1]$ that facilitate fast inference in this setting. We allow the risk distributions to vary by race and by county to accommodate the fact that the true prevalence of COVID-19 can vary by race and county. To complete the Bayesian specification, we must place priors on the latent parameters, which we describe in our full model specification, available online.\footnote{\url{https://github.com/epierson9/disease-testing-thresholds/blob/master/poisson_mixture_model_no_deltas.stan}} We infer posteriors over the latent parameters using Hamiltonian Monte Carlo~\citep{neal1994improved}, implemented in the probabilistic programming language Stan~\citep{carpenter2017stan}.

\section{Results} 

We fit the model to cumulative COVID-19 test and case count data through August 16, 2020 in the US state of Indiana, broken down by race and county (further data details in Appendix~\ref{sec:data_processing_details}). We infer testing thresholds for non-Hispanic Black, non-Hispanic white, and Hispanic populations. The primary latent parameters of interest are the inferred testing thresholds $z_{rd}$ for each race and county; we plot these in Figure \ref{fig:indiana_thresholds}. Inferred thresholds for minorities are generally higher than those for whites in the same county, suggesting that minorities are under-tested relative to whites: that is, tested only when they have a higher probability of having COVID-19. Consistent with this, the raw positivity rates (Figure \ref{fig:indiana_positivity_rate}) also show racial disparities, but they are less consistent, and this analysis is not robust to infra-marginality. (Appendix~\ref{sec:data_processing_details} includes additional model results. Figure \ref{fig:indiana_risk_distributions} plots the inferred risk distributions, illustrating that there are indeed differences across race groups. Figure \ref{fig:indiana_ppcs} plots posterior predictive checks, a standard check in Bayesian inference; Table \ref{tab:specification_robustness_checks} shows that our main results remain robust across alternate specifications.)

\begin{figure}[h!]
\floatconts
  {fig:indiana_thresholds}
  {\caption{Inferred testing thresholds $z_{rd}$. Each point represents one county; the thresholds for whites is plotted on the x-axis, and for minorities on the y-axis (Black on the left and Hispanic on the right). The dotted line denotes equal thresholds; most points are above it, indicating that minorities face higher thresholds for receiving COVID-19 tests when controlling for location. Circle size represents the number of tests in a county.}}
  {\includegraphics[width=1.2\linewidth]{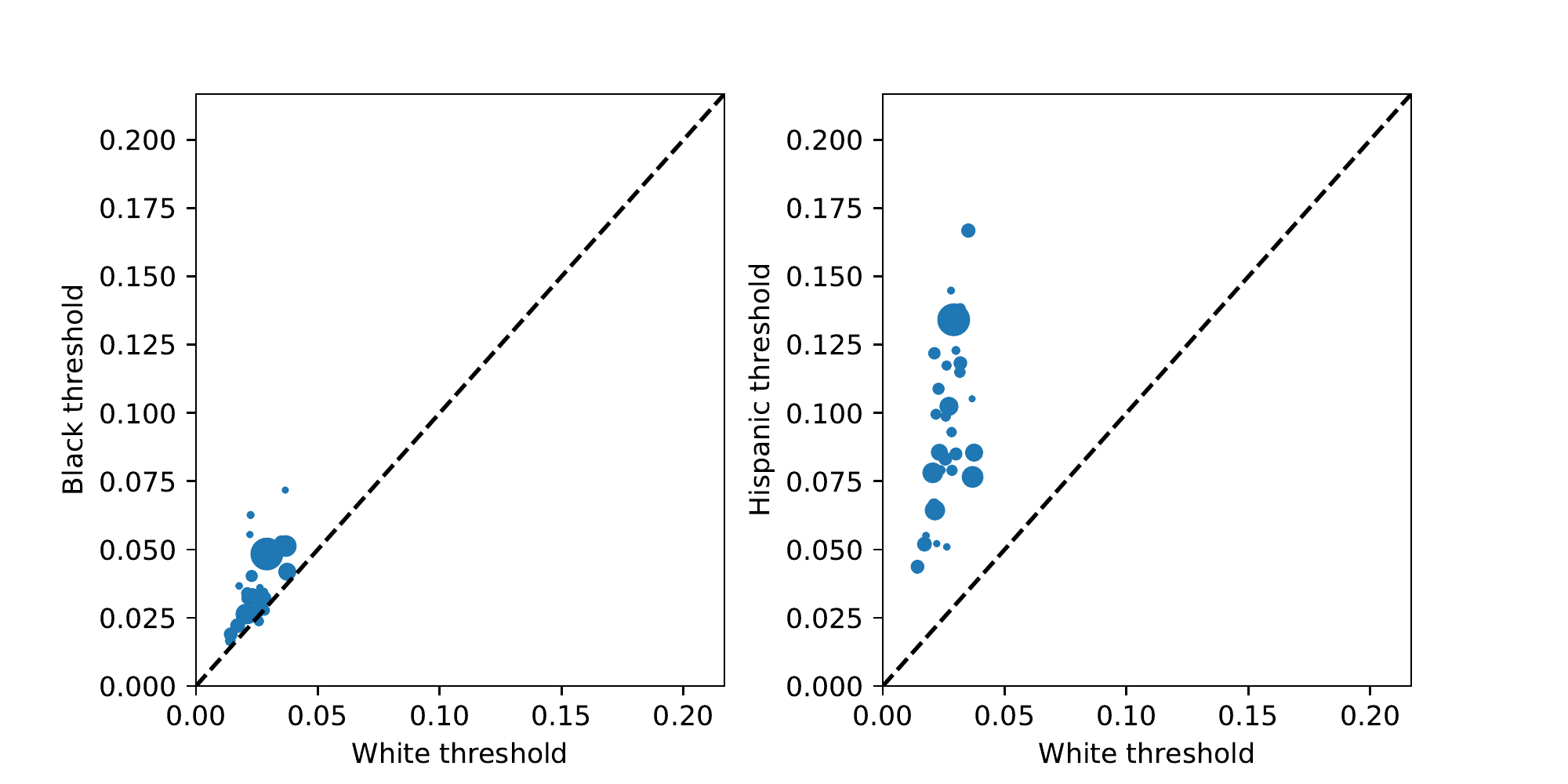}}
\end{figure}

\section{Future Directions} 

There are broader potential applications of threshold tests both in COVID-19 and in other medical conditions. While we focus here on \emph{racial} disparities in COVID-19, one could assess COVID-19 under-testing across locations or across other demographic dimensions like age. 
Beyond COVID-19, threshold tests could be used to measure racial disparities in under-testing and under-diagnosis in medicine more broadly, an issue of known concern in cardiac conditions~\cite{schulman1999effect}, attention deficit disorder~\citep{coker2016racial}, chronic obstructive lung disease~\citep{mamary2018race}, depression~\citep{sorkin2011underdiagnosed} and other psychiatric conditions~\citep{kunen2005race}. Assessing racial disparities in medical testing with threshold tests thus offers a fruitful area for further research.

\acks{Thanks to Serina Chang, Irene Chen, Sam Corbett-Davies, Pang Wei Koh, Lester Mackey, Ziad Obermeyer, Nat Roth, Leah Pierson, Miriam Pierson, Jacob Steinhardt, and seminar attendees for helpful conversations, and to Jaline Gerardin, Alexis Madrigal, and Albert Sun for data assistance.}
\bibliography{references}

\FloatBarrier
\clearpage
\appendix
\renewcommand\thefigure{\thesection\arabic{figure}}    
\renewcommand\thetable{\thesection\arabic{table}}    

\section{ } 
\label{sec:data_processing_details} 

\subsection{Data processing} 

We fit the model to cumulative COVID-19 test and case count data in Indiana through August 16, 2020 in the US state of Indiana, broken down by race and county.\footnote{We fit the model to cumulative case counts because daily case counts are not available. We note that, while the conceptual interpretation of the model makes sense even for cumulative case counts, since the sum of daily Poisson draws will itself be a Poisson draw, fitting the model to daily data (to infer time-varying thresholds and risk distributions) represents an interesting direction for future work.} We chose Indiana because it was one of the few states which made the requisite data available.\footnote{\url{https://www.coronavirus.in.gov/2393.htm}}

We infer county-specific COVID-19 testing thresholds for non-Hispanic white, non-Hispanic Black, and Hispanic populations. Indiana reports data (test and case counts) aggregated by race (eg, white or Black), and aggregated by ethnicity (Hispanic or non-Hispanic), but not data aggregated by both at once. A second caveat is that there is significant missing race/ethnicity data: the median county has ethnicity data for only about half of cases and tests, and race data for 80-90\% of cases and tests. Due to these two caveats, there are multiple potential ways of processing the raw data to produce the data we use to actually fit the model. As a sensitivity analysis, we process the data three different ways, and verify that our main conclusion (that minorities face higher testing thresholds) remains robust across all three specifications (Table \ref{tab:specification_robustness_checks}). 

\begin{enumerate} 
\item \textbf{Original specification}: We assume that race and ethnicity are independent -- eg, the fraction of whites who are Hispanic is the same as the fraction of Blacks who are Hispanic. We define the number non-Hispanic white cases as 

$$w \cdot (1 - f_{\text{hisp}}) / f_{\text{race known}}$$

where $w$ is the number of white cases in the raw data, $f_\text{hisp}$ is the proportion of cases with known ethnicity which are Hispanic, and $f_{\text{race known}}$ is the proportion of cases with known race. (All these quantities are reported in the raw data.) The definitions for tests rather than cases, and for non-Hispanic Blacks rather than non-Hispanic whites, are analogous. We define the Hispanic case count as $c \cdot f_{\text{hisp}}$, where $c$ is the total number of cases, and tests analogously. This constitutes our primary specification which we use to report our main results. 

\item \textbf{Raw counts}. We simply use the raw counts in the data for white, Black, and Hispanic cases. This method does not attempt to combine race and ethnicity or account for missing data. 
\item \textbf{Subtract ethnicity}. We use the raw counts for Black and Hispanic cases/tests, and for whites, subtract the number of Hispanic cases/tests. This method does not attempt to account for missing data. 
\end{enumerate} 

We filter for counties with Black and Hispanic populations of at least 500 to ensure that they have large enough minority populations to be able to meaningfully assess disparities. This filter retains counties containing 87\% of the Hispanic population and 98\% of the Black population. 

\subsection{Model details}
\label{sec:further_model_details} 

Our full model specification, which includes priors on all parameters and the parameterization of the thresholds and risk distributions, is available online, along all the code to reproduce our results.\footnote{\url{https://github.com/epierson9/disease-testing-thresholds}}. Here, we briefly detail how our model differs from previous threshold models. The primary difference between the original threshold model for police searches detailed in~\citet{simoiu2017problem}, and our threshold model for COVID-19 tests, is that the policing model measures disparities only among stopped drivers, whereas the COVID-19 model measures disparities in the entire population, and must therefore model and make use of population information (eg, from Census data). While the number of police searches cannot exceed the number of police stops (and the original authors therefore model the number of searches as a Binomial draw from the number of stops), the number of COVID-19 tests in a county can exceed the number of people in a location (since each person can be tested multiple times), so a Binomial model is unsuitable. ~\citet{pierson2018fast} proposes a version of the threshold test which incorporates population information, but makes use of only the proportion rather than the absolute population of each race group in each location: eg, the population information provided to their model is that ``in County X, 40\% of people are Hispanic, 20\% are white, and 40\% are Black''. This is unsuitable for our setting, because intuitively our inferences about COVID-19 testing threshold and prevalence should be very different in a county with 100 people and 10 tests, compared to a county with 10,000 people and 10 tests, even if the relative fractions of each race group remain constant. 

In the COVID-19 setting, we incorporate population information by modelling the number of tests for each race group in each county as a Poisson draw (whose rate parameter is proportional to the population of that race group in that county). Our use of a Poisson bears some similarity to the Poisson regression setting, in which a Poisson whose rate parameter depends on covariates is used to model rates in a population; a natural direction for future work is to extend our model to accommodate overdispersion via, eg, a quasi-Poisson or negative binomial model~\citep{gardner1995regression}. As a further robustness check (Table \ref{tab:specification_robustness_checks}), we ensure that our primary results remain robust when we replace the Poisson with a Binomial distribution (which is similar to the original specification in ~\citet{simoiu2017problem}), even though the latter makes less sense conceptually. 

\FloatBarrier
\setcounter{figure}{0}  
\setcounter{table}{0}    

\vspace{-10cm}

\begin{figure*}[t!]
\floatconts
  {fig:indiana_positivity_rate}
  {\caption{Positivity rates by county in Indiana. While, as with the thresholds, positivity rates are generally higher for minorities than for whites in the same county, this analysis is not robust to infra-marginality and disparities emerge slightly less consistently: for example, in 5 counties, the Black positivity rate is lower than the white positivity rate.}}
  {\includegraphics[width=0.9\linewidth]{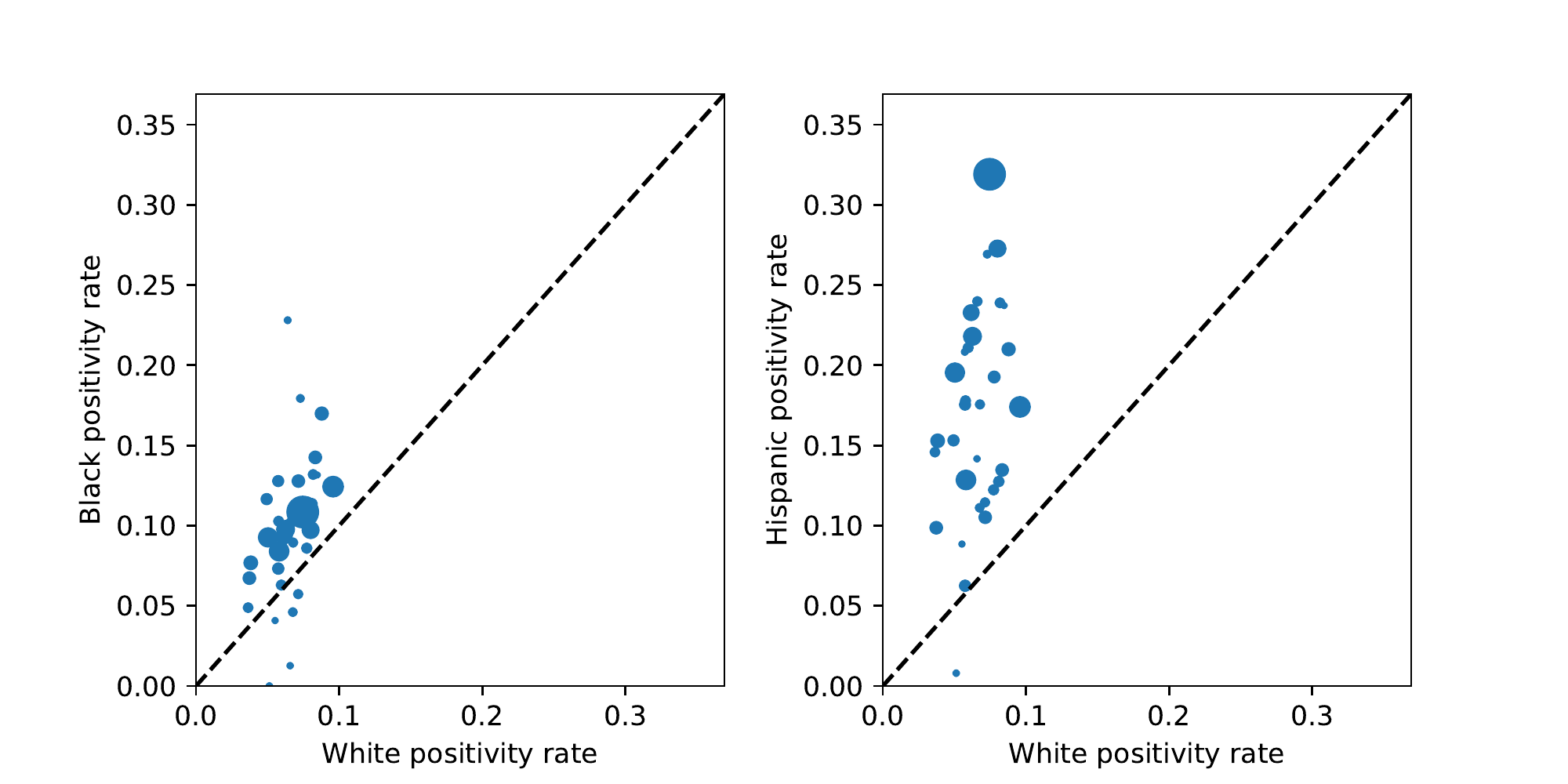}}
\end{figure*}

\begin{figure*}[b!]
\floatconts
  {fig:indiana_risk_distributions}
  {\caption{Inferred risk distributions (solid lines) and aggregated thresholds (dotted lines) by race. Both risk distributions and thresholds are aggregated across counties, weighting each county by total tests conducted in the county for people of any race. Inferred thresholds are higher for minorities, and their risk distributions are also right-shifted.}}
  {\includegraphics[width=0.9\linewidth]{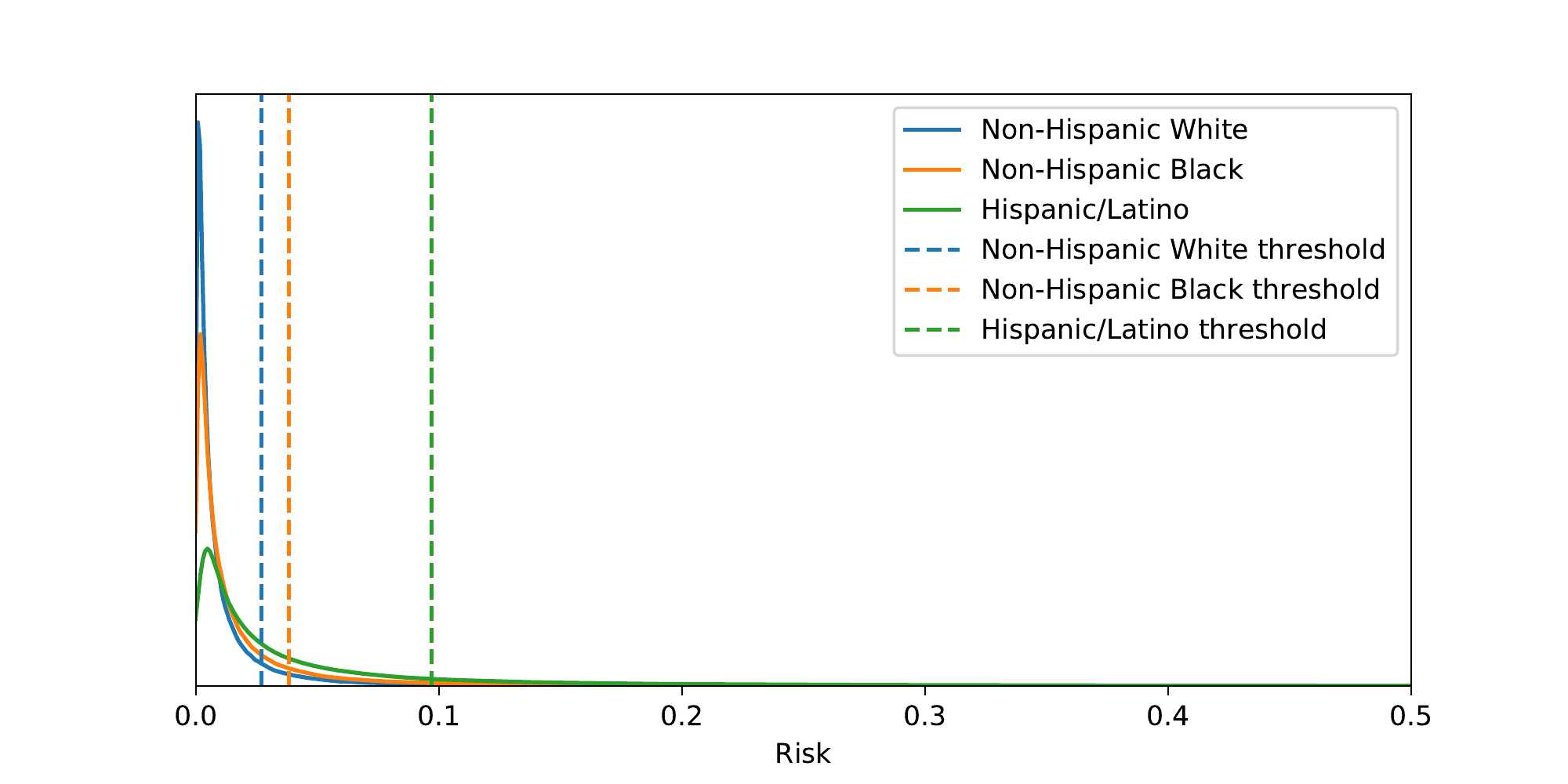}}
\end{figure*}

\begin{figure*}[htbp]
\floatconts
  {fig:indiana_ppcs}
  {\caption{Posterior predictive checks, a standard check for model fit~\citep{gelman1996posterior}. We compare observed and predicted quantities for tests per capita (top plot) and positivity rate (bottom plot). The x-axis plots the observed quantity, and the y-axis plots the error -- ie, the difference between the observed values and the model-predicted values. Points are sized proportional to the number of tests for the race group and county. The size of the point represents the number of tests in that location. Across all race groups, errors are small and there is a lack of systematic bias, validating model fit.}}
  {\includegraphics[width=.9\linewidth]{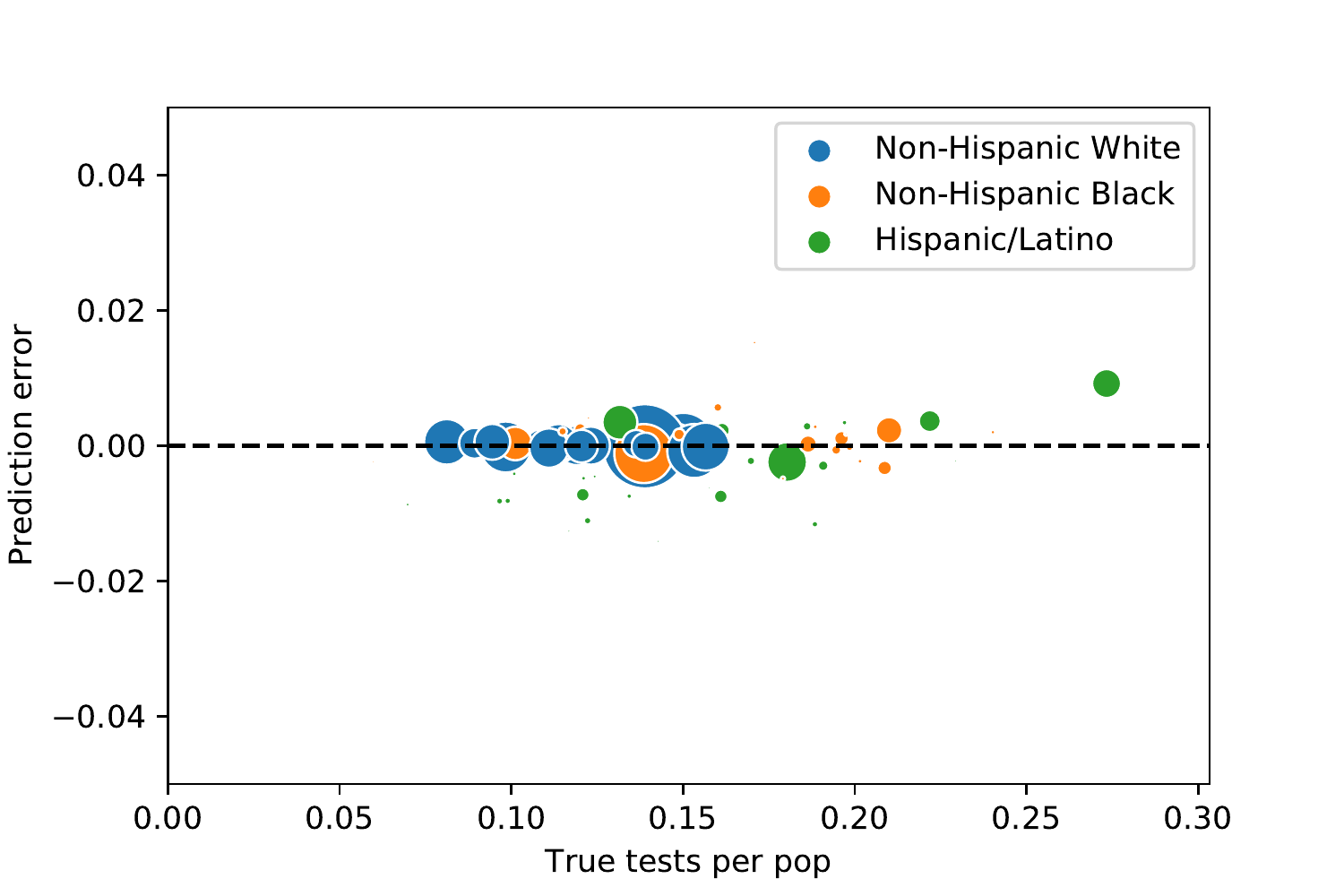}\\
  \includegraphics[width=.9\linewidth]{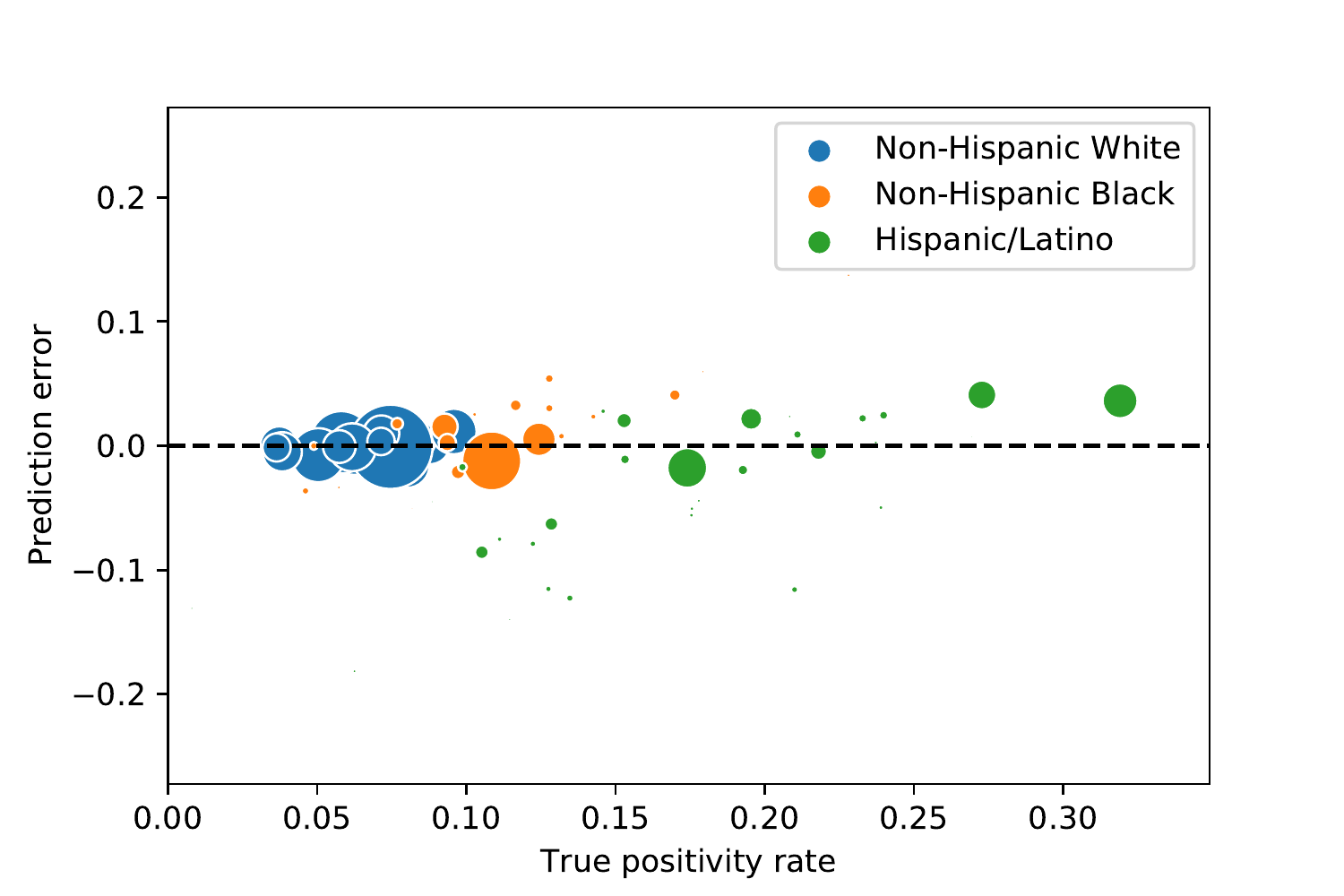}}
\end{figure*}

 \begin{table*}[htp]
\floatconts
  {tab:specification_robustness_checks}
  {\caption{Robustness checks on model and data processing. Each row reports inferred thresholds (weighted mean across counties, weighting by total tests conducted in the county) for one specification: one data processing method (first column) and one model (second column), with the primary specification which is used to generate all results in the paper in the first row. The next three columns report the inferred thresholds (reporting the mean of the posterior MCMC draws, followed by 95\% confidence interval in parentheses). The final two columns provide the ratio of minority thresholds to white thresholds. For all specifications, thresholds for Black and Hispanic populations are higher than those for whites. For more details, see Appendix \ref{sec:data_processing_details}.}}
  {
  \hspace*{-3cm}
 \small
\begin{tabular}{p{4cm}p{1.5cm}p{2.5cm}p{2.5cm}p{2.5cm}p{2.5cm}p{2.5cm}}
\toprule
Data Processing Method &     Model &    Hispanic/Latino & Non-Hispanic Black & Non-Hispanic White & Hispanic:white ratio &  Black:white ratio \\
\midrule
                         Original &   Poisson &  0.10 (0.08, 0.11) &  0.04 (0.03, 0.04) &  0.03 (0.02, 0.03) &    3.61 (3.50, 3.72) &  1.42 (1.38, 1.46) \\
              Original &  Binomial &  0.10 (0.09, 0.11) &  0.04 (0.03, 0.05) &  0.03 (0.02, 0.03) &    3.59 (3.48, 3.71) &  1.42 (1.38, 1.46) \\
    Subtract ethnicity &   Poisson &  0.17 (0.15, 0.18) &  0.07 (0.06, 0.08) &  0.04 (0.03, 0.04) &    4.35 (4.10, 4.64) &  1.82 (1.77, 1.87) \\
    Subtract ethnicity &  Binomial &  0.17 (0.15, 0.19) &  0.07 (0.06, 0.08) &  0.04 (0.03, 0.05) &    4.33 (4.08, 4.60) &  1.82 (1.78, 1.87) \\
           Raw numbers &   Poisson &  0.13 (0.11, 0.15) &  0.05 (0.04, 0.06) &  0.03 (0.03, 0.04) &    3.77 (3.51, 4.08) &  1.45 (1.41, 1.49) \\
           Raw numbers &  Binomial &  0.13 (0.11, 0.15) &  0.05 (0.04, 0.06) &  0.04 (0.03, 0.04) &    3.75 (3.50, 4.04) &  1.45 (1.41, 1.49) \\
\bottomrule
\end{tabular}
}
\end{table*}

\end{document}